\documentclass[pra,twocolumn,superscriptaddress,showkeys]{revtex4-1}

\usepackage{amsmath,amssymb,amsfonts}
\usepackage{graphicx}
\usepackage{bm}
\usepackage{hyperref}

\usepackage{braket}

\begin{document}

\title{Generalized Hund's rule for two-atom systems}

\author{Hiroki Isobe}
\affiliation{Department of Applied Physics, University of Tokyo, Tokyo 113-8656, Japan}

\author{Naoto Nagaosa}
\affiliation{RIKEN Center for Emergent Matter Science, ASI, RIKEN, Wako, Saitama 351-0198, Japan}
\affiliation{Department of Applied Physics, University of Tokyo, Tokyo 113-8656, Japan}

\date{\today}

\begin{abstract}
Hund's rule is one of the fundamentals of the correlation physics at the atomic level,
determining the ground state multiplet of the electrons. 
It consists of three laws: (i) maximum $S$ (total spin), (ii) maximum $L$ (total 
orbital angular momentum) under the constraint of (i), and (iii)
the total angular momentum $J$ is $|L-S|$ for electron number less than 
half, while $J=L+S$ for more than half due to the relativistic 
spin-orbit interaction (SOI). 
In real systems, the electrons hop between the atoms and gain the 
itinerancy, which is usually described by the band theory. 
The whole content of theories on correlation is to provide a reliable way to describe
the intermediate situation between the two limits.  
Here we propose an approach toward this goal, i.e., we study the 
two-atom systems of three $t_{2g}$ orbitals and see how the Hund's rule is modified by the
transfer integral $t$ between them. It is found that the
competition between $t$ and the Hund's coupling $J$ at each atom 
determines the crossover from the molecular orbital limit to the
strong correlation limit. 
Especially, the focus is on the generalization of the third rule, i.e., the
inter-and intra-atomic SOI's in the presence of the correlation. 
We have found that there are cases where the effective SOI's 
are appreciably enhanced by the Hund's coupling.
The conditions for the enhancement are the intermediate Hund's coupling and 
the filling of four or five electrons. 
\end{abstract}

\maketitle

\section{Introduction}

Hund's rule is one of the most important principles in the physics of strong electron correlation~\cite{*[{}] [{; }] hund1927zdd,*hund1927zdda}.
It specifies the ground state multiplet of an atom by the three conditions as follows:
(i) The total spin $S$ is maximized, because the Coulomb interaction is 
reduced by the Pauli exclusion principle for parallel spins.  (ii) The total orbital angular momentum 
$L$ is maximized within the condition (i). An intuitive explanation of this second rule is 
that the electrons with the same direction of rotation have less probability to collide with 
each other. (iii) The total angular momentum $J$ for systems with electrons less than half is
$J=|L-S|$, while that for electrons more than half is $J=L+S$. This third rule comes from the 
relativistic spin-orbit interaction (SOI) between the spin and orbital angular momenta. 

In solids, the Hund's rule and electron hopping between atoms are the two 
competing elements to determine the electronic state of correlated systems. 
When electron correlation is very strong, the electrons are almost localized at each
atom, and the local picture based on the Hund's rule is a good starting point. 
In the other limit of weak electron correlation, the band picture is the good starting point, 
where the electronic wavefunction is the extended Bloch waves, and electron correlation can be 
taken into account by the perturbation theory. 
Another possible approach to take into account the electron hopping $t$ is to consider the 
two atom systems and study the competition between $t$ and the Hund's rule coupling.
This direction has been already explored in the classical theory of Heitler and London for a 
hydrogen molecule made of two hydrogen atoms~\cite{heitler1927wna}. In this case, the two types of ground states,
i.e., (a) the singlet state made of two electrons occupying the molecular orbital,
and (b) the correlated wave function with one electron for each atom, 
are considered corresponding to the weak and strong correlation limits, respectively.
This consideration can be generalized to include the many orbitals and also the 
relativistic SOI to reveal the modification of the Hund's rule 
for the two-atom systems, which we undertake in the present paper.

Especially, the focus will be on the SOI, which is the origin of many novel 
effects such as the anomalous Hall effect~\cite{nagaosa2010ahe},  
spin Hall effect~\cite{murakami20111sh}, and topological 
insulators~\cite{*[{For recent review articles, see }] [{; }] hasan2010cti,*qi2011tia}. 
Usually strong SOI in heavy elements are required to realize these novel effects. 
For example, elements such as Bi, Hg, and Pt are main players in the physics of strong SOI. 
From the viewpoint of applications, these elements are rare and it is highly desirable 
to realize reasonably strong SOI in lighter and common elements. 
From this respect, the interplay between electron correlation and SOI has attracted recent intensive interests. 
For electrons in $d$ orbitals, the electron correlation $U$ gets stronger while the SOI weaker 
as one goes up from $5d$ to $4d$, and to $3d$ in the Periodic Table~\cite{kim2008njm}. 
In compounds consisting of $5d$ elements, both the SOI and $U$ are of the order of $0.5\ \text{eV}$, 
offering an ideal arena 
to study the interplay between these two interactions. 
For example, in Sr$_2$IrO$_4$ the $t_{2g}$ orbitals with pseudo orbital angular momentum $L_\text{eff}=1$ are coupled 
to the spin $S=1/2$ to form the effective total angular momenta $J_\text{eff}=1/2$ and $3/2$. 
Ir$^{4+}$ has $d^5$ electron configuration, and hence $J_\text{eff}=1/2$ band is half-filled. 
The width of this band is reduced by the SOI, and the reasonable $U$ is enough to localize the electron 
at each atomic site, i.e., a Mott insulator realized by the collaboration of the SOI and $U$~\cite{kim2008njm}. 
A band structure calculation proposes that the honeycomb Na$_2$IrO$_3$ 
is a weak topological insulator~\cite{shitade2009qsh}, 
while an approach from the strong coupling limit concludes that Na$_2$IrO$_3$ is a realization of the 
Kitaev spin model~\cite{jackeli2009mii}. 
A common feature of these proposals is that the interatomic SOI is essential such as the 
Rashba-type SOI~\cite{rashba1960psw} and Dzyaloshinskii--Moriya (DM) interaction~\cite{yoshida1996tm}. 
Another interesting theoretical proposal is the spontaneous symmetry breaking of the spin rotation driven 
by the Coulomb interaction, where the effective intersite SOI corresponding to the electron transfer 
with spin flip is produced~\cite{raghu2008tmi,kurita2011tif}. 
This points to an interesting possibility that the electron correlation might replace the role of SOI 
in some situations. 

In this paper, we study the Hund's rule generalized for two-atom systems with three $t_{2g}$ orbitals for each atom by exactly solving the eigenstates numerically for all the cases of electron numbers. 
We would stress that the two-site problem opens the way for many-body problems and that it brings completely different qualitative results from a single-site problem. 
In this sense, the two-site problem has a fundamental importance. 
Here we will extend the notion of the Hund's rule, examining the crossover from the weak to strong correlation, which is basically controlled by the ratio $J/t$ with $J$ ($t$) being the Hund's rule coupling for each atom (the transfer integral between the two atoms). 
The total spin $S$, orbital angular momentum $L_i$ of each atom, and the effective intra- and interatomic SOI strength are studied as functions of $J/t$. 
Through this study, we determine the condition for the enhancement of the SOI by electron correlation, and it is found that the intermediate or frustrated situations 
between the spin singlet and spin polarized states are most preferable, and based on this result candidate materials are proposed. 


\section{Model}

Let us consider a two-site model where the electron wavefunctions are bound to each atom. 
When we consider this atomic limit, the Hamiltonian is represented in the basis of atomic orbitals  
$\ket{\psi_{im\alpha}}$, where $i$ denotes an atomic site, $m$ an orbital of an atom, and $\sigma$ a spin of an electron. 
The Hamiltonian of the two-atom system is written as~\cite{imada1998mit}
\begin{equation}
\hat{H} = \sum_{i,j=1,2} \hat{H}^\text{(t)}_{ij} + \sum_{i=1,2} \hat{H}^\text{(correlation)}_i 
+ \sum_{i=1,2} \hat{H}^\text{(SO)}_i ,
\end{equation}
where
\begin{gather}
\hat{H}^\text{(t)}_{ij} = \sum_{mm'} \sum_\sigma t_{im,jm'} d^\dagger_{im\sigma} d_{jm'\sigma}, \\
\hat{H}^\text{(correlation)}_i = \sum_{m_1 m_2 m_3 m_4} \sum_{\sigma \sigma'} U_{m_1 m_2 m_3 m_4} \notag \\
	\hspace{80pt} \times d^\dagger_{im_1 \sigma}  d^\dagger_{im_2 \sigma'} d_{im_3 \sigma'} d_{im_4 \sigma}, \\
\hat{H}^\text{(SO)}_i = \sum_\alpha \sum_{mm'} \sum_{\sigma \sigma'} \zeta_{nl} d^\dagger_{im\sigma} (l_\alpha)_{mm'} (s_\alpha)_{\sigma\sigma'} d_{im'\sigma'}. 
\end{gather}
Here $d$ ($d^\dagger$) is the electron annihilation (creation) operator. 
The matrix elements of the transfer matrix $t_{im,jm'}$ are given by the Slater-Koster tables~\cite{slater1954slm}. 
$(l_\alpha)_{mm'}$ and $(s_\alpha)_{\sigma\sigma'}$ $(\alpha = x,y,z)$ are the matrix elements of the orbital and spin angular momenta, respectively. 
The parameter $\zeta_{nl}$ for the SOI depends on the principal and angular momentum quantum numbers  
($n$ and $l$, respectively)~\cite{sugano1970mtm}. 

We focus on $t_{2g}$ orbitals in the following analysis, and $m$ corresponds to $d_{yz}$, $d_{zx}$, $d_{xy}$ orbitals.
In this case, the correlation part of the atomic Hamiltonian $\hat{H}^\text{(correlation)}$ is given by the 
Kanamori Hamiltonian~\cite{kanamori1963eca}:
\begin{align}
\label{eq:Kanamori}
& \hat{H}^\text{(Kanamori)} \notag \\
= & U \sum_m \hat{n}_{m\uparrow} \hat{n}_{m\downarrow} + U' \sum_{m \neq m'} \hat{n}_{m\uparrow} \hat{n}_{m'\downarrow}
\notag \\
& +(U'-J) \sum_{m<m', \sigma} \hat{n}_{m\sigma} \hat{n}_{m'\sigma} 
 - J \sum_{m \neq m'} d^\dagger_{m\uparrow} d_{m\downarrow} d^\dagger_{m'\downarrow} d_{m'\uparrow} \notag \\
& + J \sum_{m \neq m'} d^\dagger_{m\uparrow} d^\dagger_{m\downarrow} d_{m'\downarrow} d_{m'\uparrow}.
\end{align}
A site index $i$ is omitted for simplicity in this paragraph. 
Here we define the number operator $\hat{N}$, and the orbital and angular momentum operators $\hat{\bm{L}}$, $\hat{\bm{S}}$ as follows:
$\hat{N} = \sum_{m\sigma} d^\dagger_{m\sigma} d_{m\sigma}$,
$\hat{L}_\alpha = \sum_{mm'} \sum_\sigma d^\dagger_{m\sigma} (l_\alpha)_{mm'} d_{m'\sigma}$, and 
$\hat{S}_\alpha = \sum_m \sum_{\sigma \sigma'} d^\dagger_{m\sigma} (s_\alpha)_{\sigma\sigma'} d_{m\sigma'}$,
where the matrix elements are given by 
$(l_\alpha)_{mm'} = i \epsilon_{\alpha mm'}$ and $(s_\alpha)_{\sigma\sigma'} = (\sigma_\alpha)_{\sigma\sigma'}/2$
with $\sigma_\alpha$ being a Pauli matrix.
(We set $\hbar=1$ throughout this paper.) 
We note that $d_{yz}$, $d_{zx}$, $d_{xy}$ orbitals are associated with labels $m, m' = x, y, z$, respectively. 
The Kanamori Hamiltonian is represented by using $\hat{N}$, $\hat{\bm{L}}$, and $\hat{\bm{S}}$ as 
\begin{align}
\label{eq:kanamori}
& \hat{H}^\text{(Kanamori)} \notag \\
= & \frac{1}{4} (3U' - U) \hat{N} (\hat{N} -1) + (U' - U) \hat{\bm{S}}^2 \notag \\
& + \frac{1}{2} (U' - U - J) \hat{\bm{L}}^2 + \left( \frac{7}{4}U - \frac{7}{4}U' -J \right) \hat{N} \notag \\
& + (U' - U + 2J) \sum_{m \neq m'} d^\dagger_{m\uparrow} d^\dagger_{m\downarrow} d_{m'\downarrow} d_{m'\uparrow} .
\end{align}
This Hamiltonian has $SU(2) \otimes SO(3)$ symmetry for spin and orbital degrees of freedom with $U = U' + 2J$, 
since the last term of Eq.~\eqref{eq:kanamori} vanishes~\cite{sugano1970mtm}. 
We assume that the relation $U = U' + 2J$ is always satisfied.  

We exactly diagonalize the Hamiltonian to calculate the 
following expectation values: 
the local spin angular momentum $\overline{\bm{S}_i^2} = \braket{ \psi_0| \hat{\bm{S}}_i^2 | \psi_0}$ ($i=1,2$), 
the total spin angular momentum $\overline{\bm{S}^2} = \braket{ \psi_0| (\hat{\bm{S}}_1 + \hat{\bm{S}}_2)^2 | \psi_0}$, 
the spin correlation $\overline{\bm{S}_1\cdot\bm{S}_2} = \braket{ \psi_0| \hat{\bm{S}}_1 \cdot \hat{\bm{S}}_2 | \psi_0}$, 
the local orbital angular momentum $\overline{\bm{L}_i^2} = \braket{ \psi_0| \hat{\bm{L}}_i^2 | \psi_0}$, 
the effective transfer 
\begin{equation}
\bar{t}_{mm'} = -\frac{1}{2} \braket{ \psi_0 | \sum_\sigma [ d^\dagger_{1m\sigma} d_{2m'\sigma} + \text{H.c.}] | \psi_0}, 
\end{equation}
the effective on-site SOI 
\begin{equation}
\bar{\lambda}_i = - \braket{ \psi_0 | \sum_\alpha \sum_{mm'} \sum_{\sigma \sigma'} d^\dagger_{im\sigma} (l_\alpha)_{mm'} (s_\alpha)_{\sigma\sigma'} d_{im'\sigma'} | \psi_0},
\end{equation}
and the effective spin-dependent hopping amplitude  
\begin{align}
& \bar{t}^\alpha_{\text{SO} mm'} \notag \\
= & \frac{1}{2} \braket{ \psi_0 | \sum_{\sigma\sigma'}
[ d^\dagger_{1m\sigma} (l_\alpha)_{mm'} (s_\alpha)_{\sigma\sigma'} d_{2m'\sigma'} + \text{H.c.} ] | \psi_0 }.
\end{align}
Here $\ket{\psi_0}$ represents the ground state, and if the ground states are degenerate due to the Kramers degeneracy 
for odd electron systems, for example, we will take an average over all the degenerate ground states with equal weights. 
The eigenvalues of angular momenta $\hat{\bm{S}_i^2}$, $\hat{\bm{S}^2}$, and $\hat{\bm{L}_i^2}$ 
are given by $S_i (S_i +1)$, $S(S+1)$, and $L_i (L_i +1)$, respectively. 
We generally write the expectation values 
$\overline{\bm{S}_i^2} = \bar{S}_i (\bar{S}_i +1)$, 
$\overline{\bm{S}^2} = \bar{S} (\bar{S} +1)$, and 
$\overline{\bm{L}_i^2} = \bar{L}_i (\bar{L}_i +1)$. 
$\bar{t}^\alpha_{\text{SO} mm'}$ can be regarded as the effective interatomic SOI. 
This type of parameter has appeared in the models of topological insulators, such 
as the Kane-Mele model~\cite{kane2005zto,kane2005qsh} and 
Fu-Kane-Mele model~\cite{fu2007tii}.  
We note here that $\bar{t}^\alpha_{\text{SO} mm'}$ is nonzero only when the indices are aligned in cyclic order of $x, y, z$ 
because of the matrix element $(l_\alpha)_{mm'} = i\epsilon_{\alpha mm'}$. 

\begin{figure} 
\centering
\includegraphics[width=\hsize]{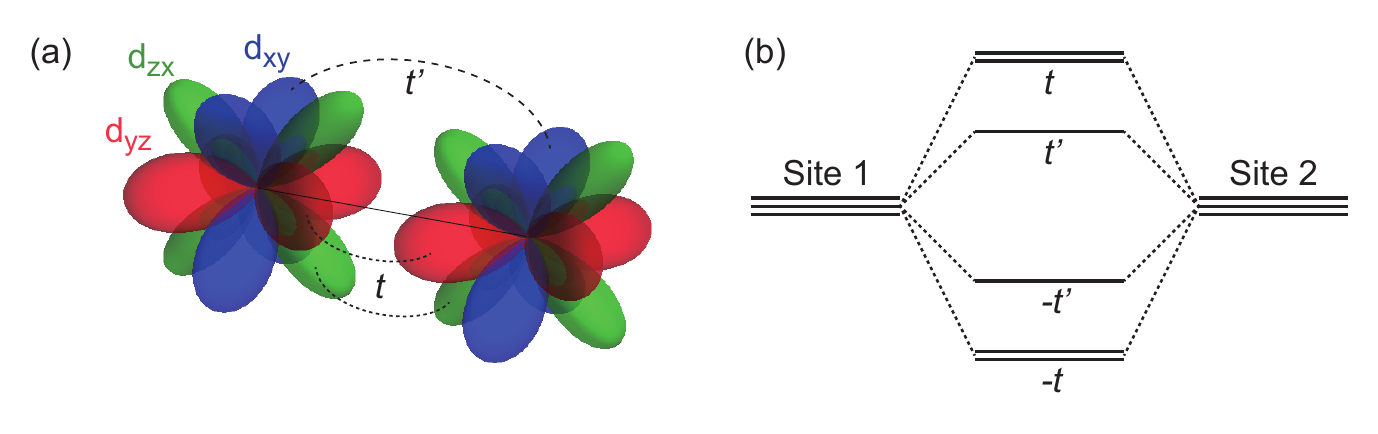}
\caption{(Color online) 
(a) Two-site model of $t_{2g}$ orbitals. Two atoms are alined along the $z$ axis. The $d_{yz}$ and $d_{zx}$ orbitals form $\pi$ bonds 
with the transfer integral $t$ and the $d_{xy}$ orbitals form $\delta$ bond with the transfer integral $t'$. 
(b) Schematic picture of the molecular orbitals. The bonding orbitals of $d_{yz}$ and $d_{zx}$ are more stable than that of $d_{xy}$ 
because of the larger overlap between the two sites. 
}
\label{fig:model}
\end{figure}

We choose the $z$ direction parallel to the bond between the two atoms and the transfer matrix as 
$t_{yz,yz} = t_{zx,zx} = t$ for the $\pi$ bonds and $t_{xy,xy} = -t'$ for the $\delta$ bond.
According to the bonding of the two atoms, the symmetry of the $t_{2g}$ orbitals is reduced from $SO(3)$ to $SO(2)$. 
We set $t'=0.5t$ and the strength of the SOI $\zeta = 0.1t$ in the following analysis. 
The model we consider here is illustrated in Fig.~\ref{fig:model}. 
Since we have two sites with electron hopping, there is a formation of molecular orbitals. 
We will examine below the fate of the Hund's rule in this two-site model.  

The number of electrons in this two-atom model ranges from zero to 12, and 
the $n$-electron system ($n \leq 6$) is complementary to the $(12-n)$-electron system 
since we consider only $t_{2g}$ orbitals. 
Therefore, we will analyze the case with one to six electrons.

\section{Results}

\subsection{One electron}
There is no electron correlation effect in one-electron systems, and thus the expectation values do not have 
dependence on $U$, $U'$, and $J$. 
The results are given as follows: 
the local spin angular momentum $\overline{\bm{S}_i^2}=0.375$ for site $i(=1,2)$; 
the total spin angular momentum $\overline{\bm{S}^2}=0.375$; 
the spin correlation $\overline{\bm{S}_1 \cdot \bm{S}_2} = 0$; 
the local orbital angular momentum $\overline{\bm{L}_i^2}=1$; 
the effective transfer $\bar{t}_{yz,yz}=\bar{t}_{zx,zx}=0.25$, $\bar{t}_{xy,xy}=0$; 
the effective interatomic SOI $\bar{t}^z_{\text{SO} yz,zx} =0.125$, $\bar{t}^x_{\text{SO} zx,xy} = \bar{t}^y_{\text{SO} xy,yz} = 0$; 
the effective on-site SOI $\bar{\lambda} = 0.25$.

\subsection{Two and three electrons}

\begin{figure} 
\centering
\includegraphics[width=\hsize]{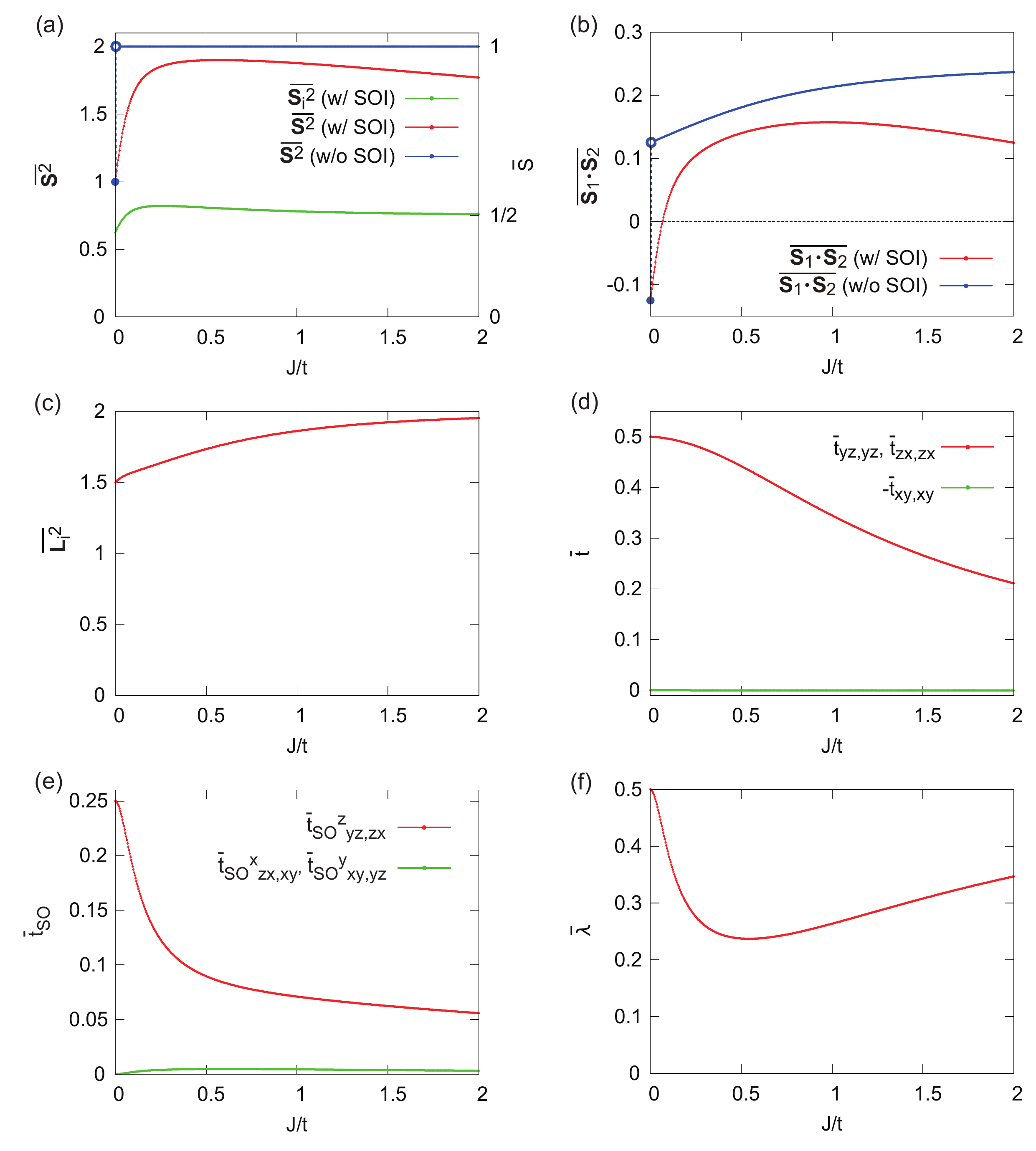}
\caption{(Color online) 
Results of exact diagonalization for the two-electron system. 
We set $U'=5J$ with the condition $U=U'+2J$. 
(a) Local spin angular momentum $\overline{\bm{S}_i^2} = \bar{S}_i (\bar{S}_i +1)$ ($i=1,2$) in the presence of SOI (green) and total spin angular momentum $\overline{\bm{S}^2} = \bar{S} (\bar{S} +1)$ in the presence (red) and absence (blue) of SOI.  
The right vertical axis measures the magnitude of spin, or spin quantum number, i.e., $\bar{S}_i$ or $\bar{S}$. 
The ground states at $J=0$ without SOI consist of three $S=0$ states and a triplet $S=1$, and for $J \neq 0$ it consists only of $S=1$. 
(b) Spin correlation $\overline{\bm{S}_1 \cdot \bm{S}_2}$ in the presence (red) and absence (blue) of SOI. 
(c) Local orbital angular momentum $\overline{\bm{L}_i^2}$ ($i=1,2$). 
Values in the graphs are those in the presence of SOI unless noted explicitly. 
(d) Effective transfers $\bar{t}$\,'s. $\bar{t}_{xy, xy}$ has very small values for all range of $J$ and is of the order of $10^{-4}$ 
because the $d_{xy}$ orbital is almost empty. 
(e) Effective interatomic SOI $\bar{t}_\text{SO}$'s. Even though $\bar{t}^x_{\text{SO} zx,xy}$ and $\bar{t}^y_{\text{SO} xy,yz}$ are 
small and have values around 0.01, the ratio $\bar{t}_\text{SO}/\bar{t}$ can have a large value due to the even smaller $\bar{t}_{xy,xy}$. 
(f) Effective on-site SOI $\bar{\lambda}$. 
We can find the rapid change of some values near $J/t=0$. 
It is caused by the change in the ground state due to the electron correlation effect. Its width is characterized by the strength of the SOI $\zeta$.
}
\label{fig:2e}
\end{figure}

When we choose $J/t$ and $U'/t$ as the free parameters with $U= U'+2J$, it is found that the expectation values depend mainly on $J/t$. 
It seems natural since at each site $\hat{\bm{S}_i^2}$ and $\hat{\bm{L}_i^2}$ are subject only to the Hund's coupling $J$, as we can confirm it from the Kanamori Hamiltonian \eqref{eq:Kanamori}. 
Therefore, we will show results of the $J/t$ dependence hereafter fixing $U'$, and below we present the results for $U'=5J$. 
For the two-electron system, the results of exact diagonalization are given in Fig.~\ref{fig:2e}.

\begin{figure} 
\centering
\includegraphics[width=\hsize]{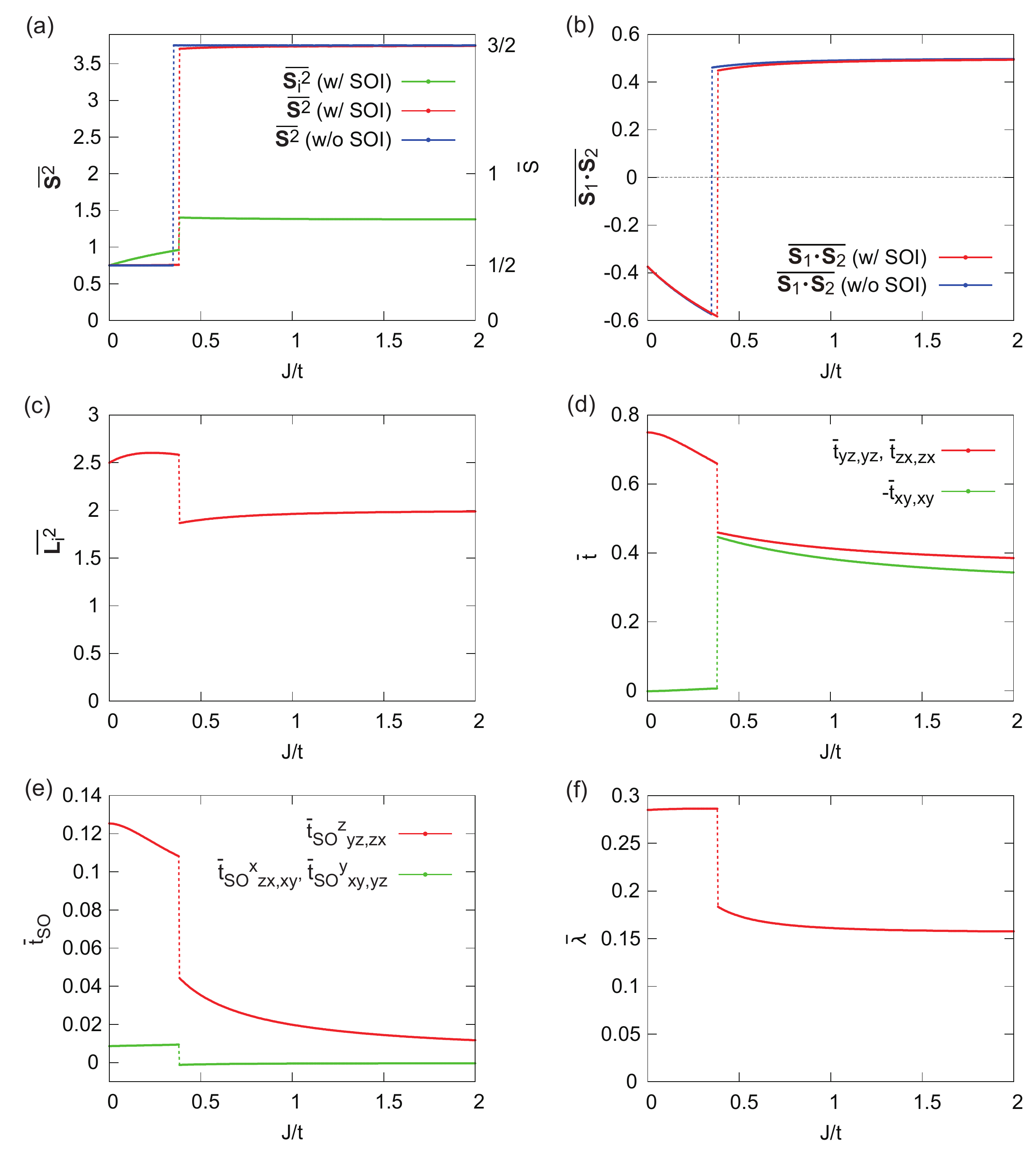}
\caption{(Color online) 
Results of exact diagonalization for the three-electron system. 
(a) Local spin angular momentum $\overline{\bm{S}_i^2}$ ($i=1,2$) in the presence of SOI (green) and total spin angular momentum $\overline{\bm{S}^2}$ in the presence (red) and absence (blue) of SOI. 
(b) Spin correlation $\overline{\bm{S}_1 \cdot \bm{S}_2}$ in the presence (red) and absence (blue) of SOI. 
The antiferromagnetic spin correlation changes to the ferromagnetic correlation by increasing $J$, accompanying the transition 
from the mostly low-spin state to the mostly high-spin state. 
(c) Local orbital angular momentum $\overline{\bm{L}_i^2}$ ($i=1,2$). 
(d) Effective transfer $\bar{t}$'s and (e) effective interatomic SOI $\bar{t}_\text{SO}$'s. 
Small values of green lines in (d) and (e) for $J/t \lesssim 0.35$ come from the small occupation of the $d_{xy}$ orbitals. 
(f) Effective on-site SOI $\bar{\lambda}$. 
These figures exhibit a discontinuity around $J/t \approx 0.35$ and the range of $J/t$ can be separated into two regions: 
the low-spin state ($S=1/2$) and the high-spin state ($S=3/2$). 
$\bar{t}^z_{\text{SO} yz,zx}$ and $\bar{\lambda}$ are relatively large for the low-spin state.
}
\label{fig:3e}
\end{figure}

Figure \ref{fig:2e}(a) shows the local and total spin angular momenta $\overline{\bm{S}_i^2} = \bar{S}_i (\bar{S}_i +1)$ ($i=1,2$) and $\overline{\bm{S}^2} = \bar{S} (\bar{S} +1)$, respectively. 
Now two sites are equivalent, so that the local spin angular momenta have the same values, $\overline{\bm{S}_1^2} = \overline{\bm{S}_2^2}$. 
The right vertical axis measures $\overline{\bm{S}_i^2}$ and $\overline{\bm{S}^2}$ in terms of their magnitude, namely $\bar{S}_i$ and $\bar{S}$. 
The nature of the ground states for different $J/t$ can be understood by focusing on the total spin quantum number $\bar{S}$, 
though it is not a good quantum number in the presence of SOI. 
In the absence of SOI, the ground states at $J=0$ consist of three singlets ($S=0$) and one triplet ($S=1$), i.e., three states of $S=1$.  
On the other hand, for $J \neq 0$, the ground states consist only of $S=1$ states. 
Namely, an infinitesimal Hund's coupling $J$ lifts the degeneracy at $J=+0$ by lowering the energy of the $S=1$ states. 
For the two-electron system around $J=0$, perturbative treatment of SOI is not allowed because the ground states are not adiabatically connected at $J=+0$. 
In other cases we will see below, the ground state without SOI is specified by a single value of $S$, and small SOI can be treated perturbatively. 
According to degenerate perturbation theory, the lowest order correction of the ground state including SOI is constructed by a linear combination 
of the degenerate ground states without SOI. 

Then, we consider the spin correlation $\overline{\bm{S}_1 \cdot \bm{S}_2}$ [Fig.~\ref{fig:2e}(b)]. 
We note that the spin correlation is obtained by the difference between the total and local spin angular momentum: 
\begin{equation}
\label{eq:spin_cor}
\overline{\bm{S}_1 \cdot \bm{S}_2} = \frac{1}{2}\overline{\bm{S}^2} - \overline{\bm{S}_i^2}. 
\end{equation}
For the two-electron system, the singlet configuration ($S=0$) gives the expectation value $\overline{\bm{S}_1 \cdot \bm{S}_2} = -3/4$, 
while the triplet ($S=1$) $\overline{\bm{S}_1 \cdot \bm{S}_2} = 1/4$. 
Recalling that the ground states at $J=0$ consist of three $S=0$ states and three $S=1$ states and that the Hund's coupling lowers the energy of $S=1$ states, 
we understand the behavior of $\overline{\bm{S}_1 \cdot \bm{S}_2}$ that the ground state at $J=0$ shows antiferromagnetic correlation 
and that it turns ferromagnetic with small $J$ of the order of SOI. 
There occurs the discontinuous jump of the spin correlation at $J=0$ without SOI. 

The behavior of the local orbital angular momentum $\overline{\bm{L}_i^2}$ is shown in Fig.~\ref{fig:2e}(c). 
For the two-electron system, it increases monotonically as $J/t$. 

The effective transfers $\bar{t}$'s are presented in Fig.~\ref{fig:2e}(d). 
There are two different values of $\bar{t}$'s, which originate from the reduction of the symmetry of the $t_{2g}$ orbitals from $SO(3)$ to $SO(2)$.  
Here $\bar{t}_{xy,xy}$ has tiny values and is of the order of $10^{-4}$, since the relevant orbitals for the two-electron system are $d_{yz}$ and $d_{zx}$ orbitals 
(or equivalently $\pi$ bonding orbitals) and $d_{xy}$ orbitals are almost empty. 
Another point is that the effective transfers become smaller as the electron correlation is increased. 
This loss of itinerancy indicates that electrons tend to localize. 

The SOI mixes high-spin and low-spin states, and induces the effective interatomic SOI $\bar{t}_\text{SO}$ [Fig.~\ref{fig:2e}(e)] 
as well as the effective on-site SOI $\bar{\lambda}$ [Fig.~\ref{fig:2e}(f)]. 
In the case of two electrons, there is a rough correspondence between  $\bar{t}_\text{SO}$ and the product of $\bar{t}$ and $\bar{\lambda}$. 
However, this rule does not apply in several cases discussed later. 
We note that the ratio $\bar{t}_\text{SO}/\bar{t}$ measures the twist of spins and orbitals between the two atoms.

The results of the three-electron system are shown in Fig.~\ref{fig:3e}. 
We can find the discontinuity in the expectation values at $J/t \approx 0.35$, and it shows a sign of a change in the ground states. 
This jump indicates that the Hamiltonian matrix is block diagonal due to some symmetries and consequent level crossing, as analyzed in the Appendix \ref{sec:symmetry}. 
Actually, in the case of two electrons, the spin-orbit rotational symmetry $\mathcal{SO}$ is broken by the SOI and consequently the discontinuity at $J=0$ is lifted. 
In the case of three electrons, on the other hand, the discontinuity comes from both $\mathcal{SO}$ and the exchange symmetry $\mathcal{X}$ of the two sites. 
Therefore, even though $\mathcal{SO}$ is broken by the SOI, the discontinuity still remains. 

To gain the transfer energy, low-spin states are favored, whereas high-spin states are more stable according to the Hund's rule. 
Now the ground state is a low-spin state in the small-$J/t$ region (approximately $S=1/2$), and 
it becomes a high-spin state in the large-$J/t$ region (approximately $S=3/2$). 
Usually the competition between the low-spin and high-spin states in $d$ orbitals is discussed from the viewpoint of the one 
between the Hund's coupling $J$ and the crystal field splitting $\Delta$ between $t_{2g}$ and $e_g$ orbitals.  
In the present situation, however, we consider the splitting among the bonding and antibonding orbitals of the $\pi$ and $\delta$ bonds, 
which competes with the Hund's coupling. 
When the Hund's coupling $J$ is relatively strong compared to the transfer energy $t$, the system obeys the Hund's rule instead of the energy gain by electron hopping.

As for the spin correlation $\overline{\bm{S}_1 \cdot \bm{S}_2}$, the ground states for the small-$J/t$ region are antiferromagnetic, and 
it turns out to be ferromagnetic for the large-$J/t$ region. 
The evidence of the $S=3/2$ state is also implied in $\bar{t}_{xy,xy}$. 
The relatively large magnitude of $\bar{t}_{xy,xy}$ in the large-$J/t$ region implies that the $d_{xy}$ orbital is also partially occupied, and 
the ferromagnetic correlation means that the electron spins are aligned to make $S=3/2$ state by using all three $t_{2g}$ orbitals.  
The behaviors of the effective SOI's and the effective transfers are similar to those of the two-electron system except the discontinuity discussed above. 
The effective SOI's are also largest in the low-spin states, i.e., the small-$J/t$ region. 
This means that there is no enhancement of the effective SOI's by electron correlation for the case of two and three electrons.

\subsection{Four and five electrons}

\begin{figure*} 
\centering
\includegraphics[width=0.85\hsize]{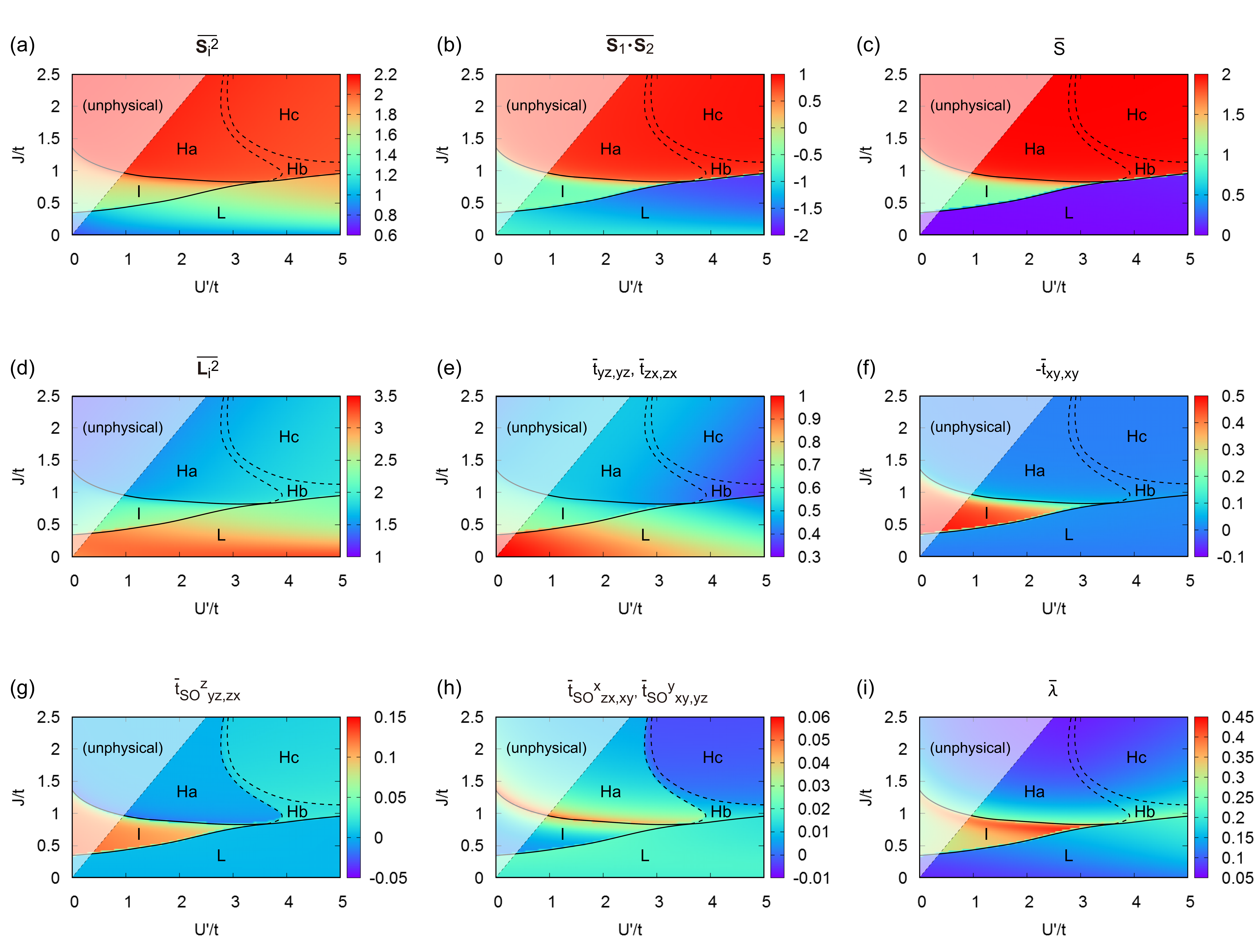}
\caption{(Color online) 
Results of exact diagonalization for the four-electron system. 
These show rather complex dependence on $U'$ and $J'$, and we thus show the whole maps by changing $U'$ and $J$ with the constraint $U = U' + 2J$. 
There are five distinct regions, i.e., the low-spin region (L), the intermediate-spin region (I), and three types of high-spin regions (Ha, Hb, Hc). 
In the shaded areas of $U'<J$, the interaction energy becomes effectively negative, which is unphysical~\cite{georges2013scf}. 
(a) Local spin angular momentum $\overline{\bm{S}_i^2}$, 
(b) spin correlation $\overline{\bm{S}_1 \cdot \bm{S}_2}$, 
(c) magnitude of total spin $\bar{S}$, 
(d) local orbital angular momentum $\overline{\bm{L}_i^2}$, 
(e) effective transfer $\bar{t}_{yz,yz}$, $\bar{t}_{zx,zx}$, (f) $-\bar{t}_{xy,xy}$, 
(g) effective interatomic SOI $\bar{t}^z_{\text{SO} yz,zx}$, (h) $\bar{t}^x_{\text{SO} zx,xy}$, $\bar{t}^y_{\text{SO} xy,yz}$, and 
(i) effective on-site SOI $\bar{\lambda}$. 
}
\label{fig:4e}
\end{figure*}

\begin{figure} 
\centering
\includegraphics[width=\hsize]{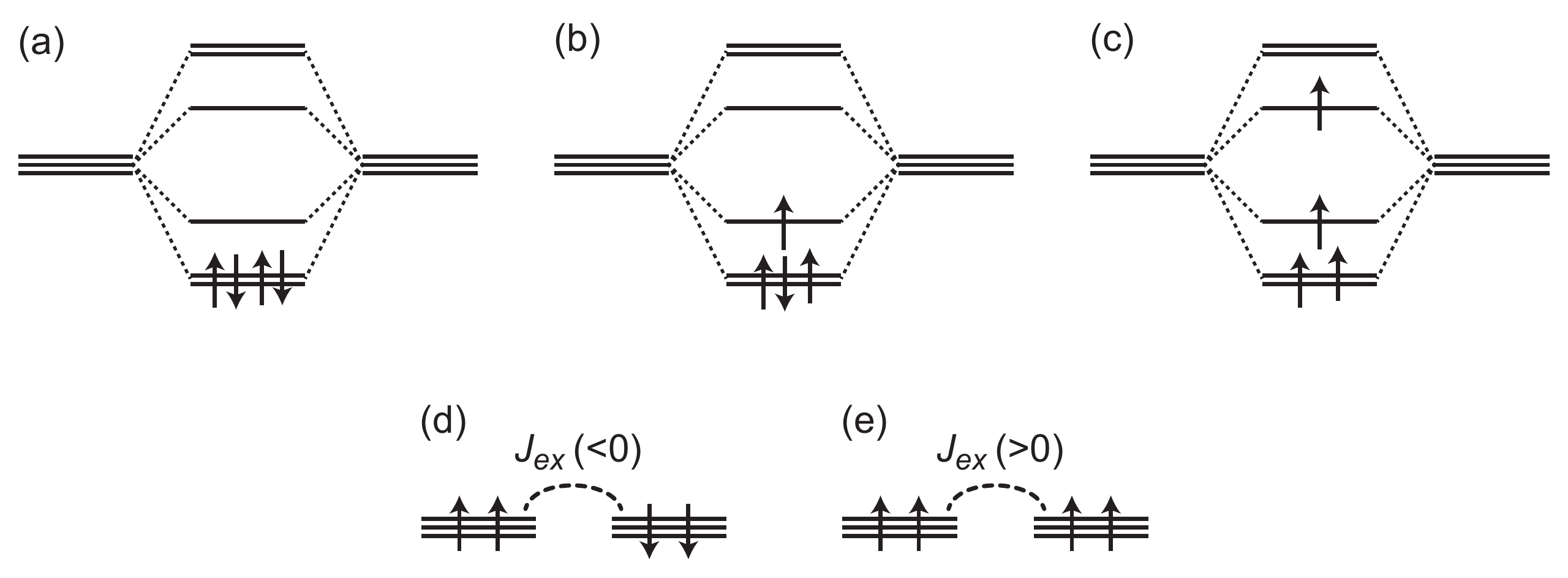}
\caption{
Schematic pictures of the ground states of the four-electron system. 
The figures in the upper row depict (a) the low-spin state $(S=0)$, (b) the intermediate-spin state $(S=1)$, and (c) the high-spin state $(S=2)$. 
In the transition from (a) to (c), a spin flip changes the total spin by one at each step.  
These pictures are valid for the relatively weak electron correlation where the description by molecular orbitals is reasonable. 
In the lower row, the ground states in the strong coupling limit (i.e., large $U$ and $U'$ limit) are presented: (d) low-spin state for small $J/t$ and (e) high-spin state for large $J/t$. 
}
\label{fig:schematic}
\end{figure}

\begin{figure} 
\centering
\includegraphics[width=\hsize]{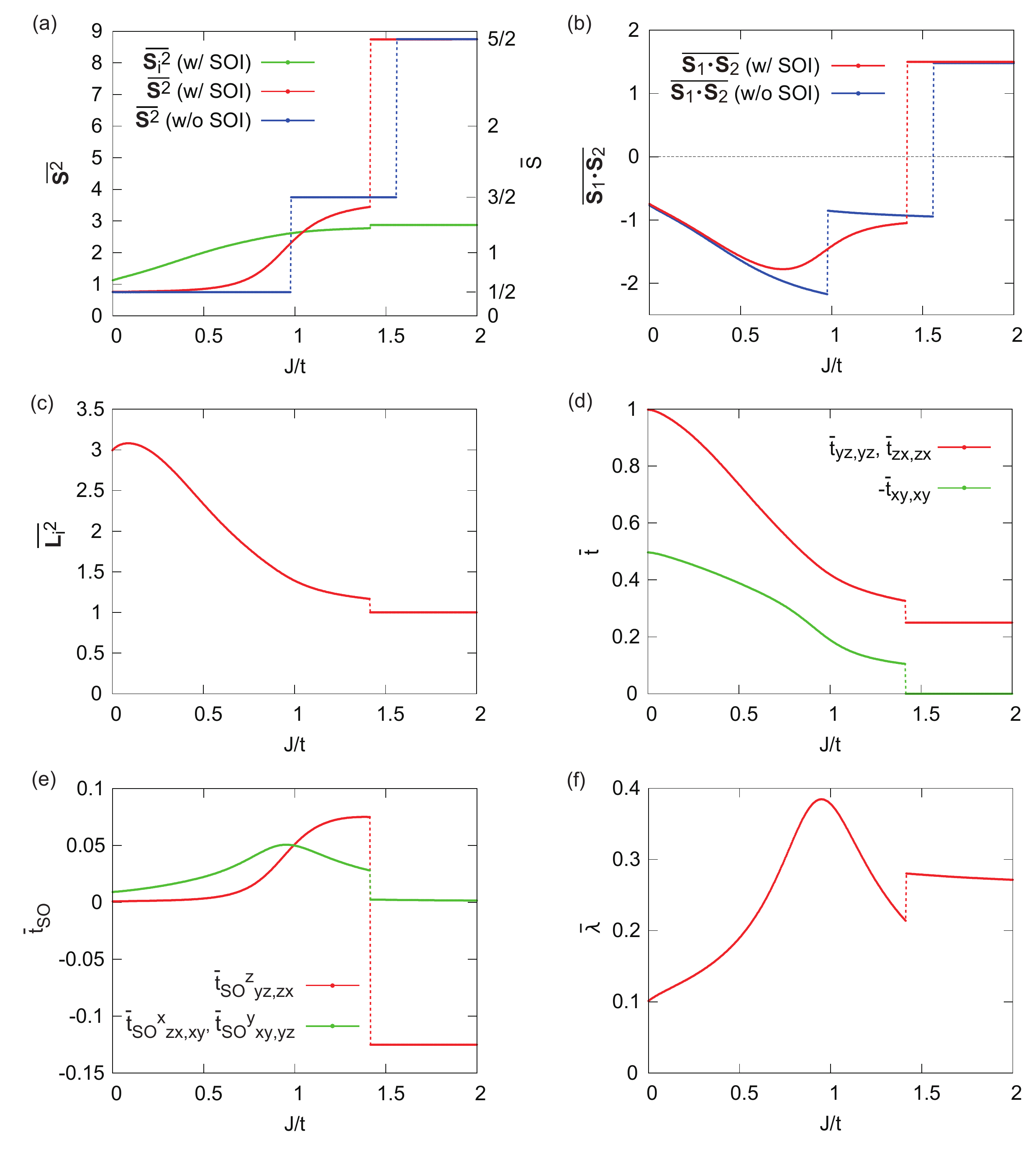}
\caption{(Color online) 
Results of exact diagonalization for the five-electron system: 
(a) local spin angular momentum $\overline{\bm{S}_i^2}$ in the presence of SOI (green) and total spin angular momentum $\overline{\bm{S}^2}$ in the presence (red) and absence (blue) of SOI, 
(b) spin correlation $\overline{\bm{S}_1 \cdot \bm{S}_2}$ in the presence (red) and absence (blue) of SOI,
(c) local orbital angular momentum $\overline{\bm{L}_i^2}$, 
(d) effective transfer $\bar{t}$'s, 
(e) effective interatomic SOI $\bar{t}_\text{SO}$'s, and 
(f) effective on-site SOI $\bar{\lambda}$. 
Apparently, there is only one discontinuity for the system with SOI, but it is because the SOI smears one of the two discontinuities. 
If the SOI is switched off, we can find two discontinuities and can identify the low-spin ($S=1/2$), intermediate-spin ($S=3/2$), and high-spin ($S=5/2$) states. 
The five-electron system has large values of $\bar{t}_\text{SO}$ and $\bar{\lambda}$ for the intermediate-spin state in common with the four-electron system.
}
\label{fig:5e}
\end{figure}

In the previous sections, we have studied the systems with two and three electrons, and found the total spin quantum number $S$ switches 
at a single value of $J/t$. For the systems with four and five electrons, this situation drastically changes. 
These systems allow the maximum spin $S=2$ or $S=5/2$, so the change of $S$ can occur twice, and the results show 
that it is actually realized (Figs.~\ref{fig:4e} and \ref{fig:5e}). 

The situation becomes complex especially for the four-electron system, since the ground state depends on the interorbital repulsion $U'$ in addition to the Hund's coupling $J$. 
As shown in Fig.~\ref{fig:4e}(c), the ground state can be separated basically into three regions: the low-spin region (L), the intermediate-spin region (I), and the high-spin region (H). 
Note that the shaded areas of each graph are unphysical since the effective Hubbard interaction $U_\text{eff}$ is reduced by $J$ to be 
$U_\text{eff} = U - 3J = U' -J$ \cite{georges2013scf} and it is negative in those areas. 
We also note that the high-spin region is further classified into three parts (Ha, Hb, and Hc in Fig.~\ref{fig:4e}) from the discontinuity in the expectation values. 
Although there is no Kramers degeneracy in a system with even number of electrons, the lowest energy state is degenerate in Ha and nearly degenerate in Hb with the energy separation less than $10^{-5}t$. 
The high-spin region is segmented due to the SOI, and the difference among Ha, Hb, and Hc appears in the effective interatomic SOI $\bar{t}_\text{SO}$ [Figs.~\ref{fig:4e}(g) and \ref{fig:4e}(h)]. 
Without SOI, the high-spin region becomes a single phase with fivefold degeneracy.

The separation of the ground state depending on $U'/t$ and $J/t$ is explained by using the schematic picture of the ground state, shown in Fig.~\ref{fig:schematic}. 
When the interorbital repulsion $U'$ and simultaneously the Hubbard $U$ are weak compared to the transfer $t$, the system is well described by the picture of molecular orbitals. 
Then if the Hund's coupling is also small, the system is stabilized by the electron transfer and thus four electrons form two singlet couplings using the $\pi$ bonding orbitals [Fig.~\ref{fig:schematic}(a)]. 
It leads to the low-spin state with the total spin $S=0$. 
As the Hund's coupling increases, a larger spin state is favored. 
If the energy gain from the Hund's coupling exceeds the difference of the transfer energies of $\pi$ and $\delta$ orbitals, namely $t-t'$, the system experiences a discontinuous transition to the intermediate-spin state [Fig.~\ref{fig:schematic}(b)]. 
Finally with the even stronger Hund's coupling it reaches the high-spin state by gaining the energy from the Hund's coupling instead of the loss of electron itinerancy [Fig.~\ref{fig:schematic}(c)]. 
We note that an itinerant electron in the $\delta$ bonding orbital, or $d_{xy}$ orbitals, exists only in the intermediate-spin region, which is clearly seen by $-\bar{t}_{xy,xy}$ [Fig.~\ref{fig:4e}(f)]. 

On the other hand, in the strong $U'$ (and also $U$) region, the isolated atom description is relevant with the exchange interaction $J_\text{ex}$ induced by the electron hopping and electron correlation. 
Since the on-site electron repulsion is strong, the electron hopping between the two sites is suppressed, and the molecular orbital description is not appropriate. 
The remnant electron hopping results in the exchange interaction between the two sites, the notion of which is in relation to the Goodenough-Kanamori rule for the superexchange coupling~\cite{goodenough1958imp,kanamori1959sia}. 
Each site occupies two electrons in this case, and the Hund's coupling $J$ makes spin 1 at each site. 
The sign of the induced exchange coupling $J_\text{ex}$ is determined in the competition between the Hund's coupling $J$ and the electron transfer $t$. 
The transfer $t$ promotes the singlet formation while $J$ prefers the larger spin; namely, $J_\text{ex}$ is negative for small $J/t$ while it becomes positive for large $J/t$ [Figs.~\ref{fig:schematic}(d) and \ref{fig:schematic}(e)].
The intermediate-spin state does not exist in the strong $U'/t$ region. 
The sign change of $J_\text{ex}$ is confirmed in the spin correlation $\overline{\bm{S}_1 \cdot \bm{S}_2}$ [Fig.~\ref{fig:4e}(b)]. 
We would add that this complex behavior in the four-electron system might have a relation to the bad metallic behavior found in the three-orbital model occupying two electrons per atom~\cite{medici2011jfi}.

The behavior of the five-electron system is rather insensitive to $U'$ and can again be explained simply by focusing on $J/t$. 
It has two discontinuous transitions in the absence of SOI, while it displays only one discontinuity in the presence of SOI (Fig.~\ref{fig:5e}).
In the first region for $J/t \lesssim 0.9$, the low-spin state ($S=1/2$) is dominant in the ground state, and as the Hund's coupling becomes strong, 
the intermediate-spin state ($S=3/2$) mainly occupies the ground state for the second region ($0.9 \lesssim J/t \lesssim 1.4$). 
These two regions are continuously connected due to the presence of SOI, i.e., broken $\mathcal{SO}$ symmetry. 
Finally, the third region ($J/t \gtrsim 1.4$) consists of the high-spin state ($S=5/2$), and it is totally ferromagnetic.

The intriguing result of the four-electron and five-electron systems is that both the effective interatomic SOI $\bar{t}_\text{SO}$ 
and the effective on-site SOI $\bar{\lambda}$ become larger by the existence of electron correlation. 
They have the largest values for the intermediate-spin states. 
This is a sharp contrast compared to the results of two-electron and three-electron systems, where the largest values of $\bar{t}_\text{SO}$ and $\bar{\lambda}$ occur at $J/t=0$. 
In the intermediate-spin region, both the spin and orbital angular momenta have moderate values, and therefore their product and consequently the effective SOI's are largest in this region. 
We also note that the absolute values of $\bar{t}^z_{\text{SO} yz,zx}$ and $\bar{t}^x_{\text{SO} zx,xy}$ (or $\bar{t}^y_{\text{SO} xy,yz}$) become comparable. 
This can be explained by an increasing occupation number of the $d_{xy}$ orbital for systems of four or more electrons.

\subsection{Six electrons}

\begin{figure} 
\centering
\includegraphics[width=\hsize]{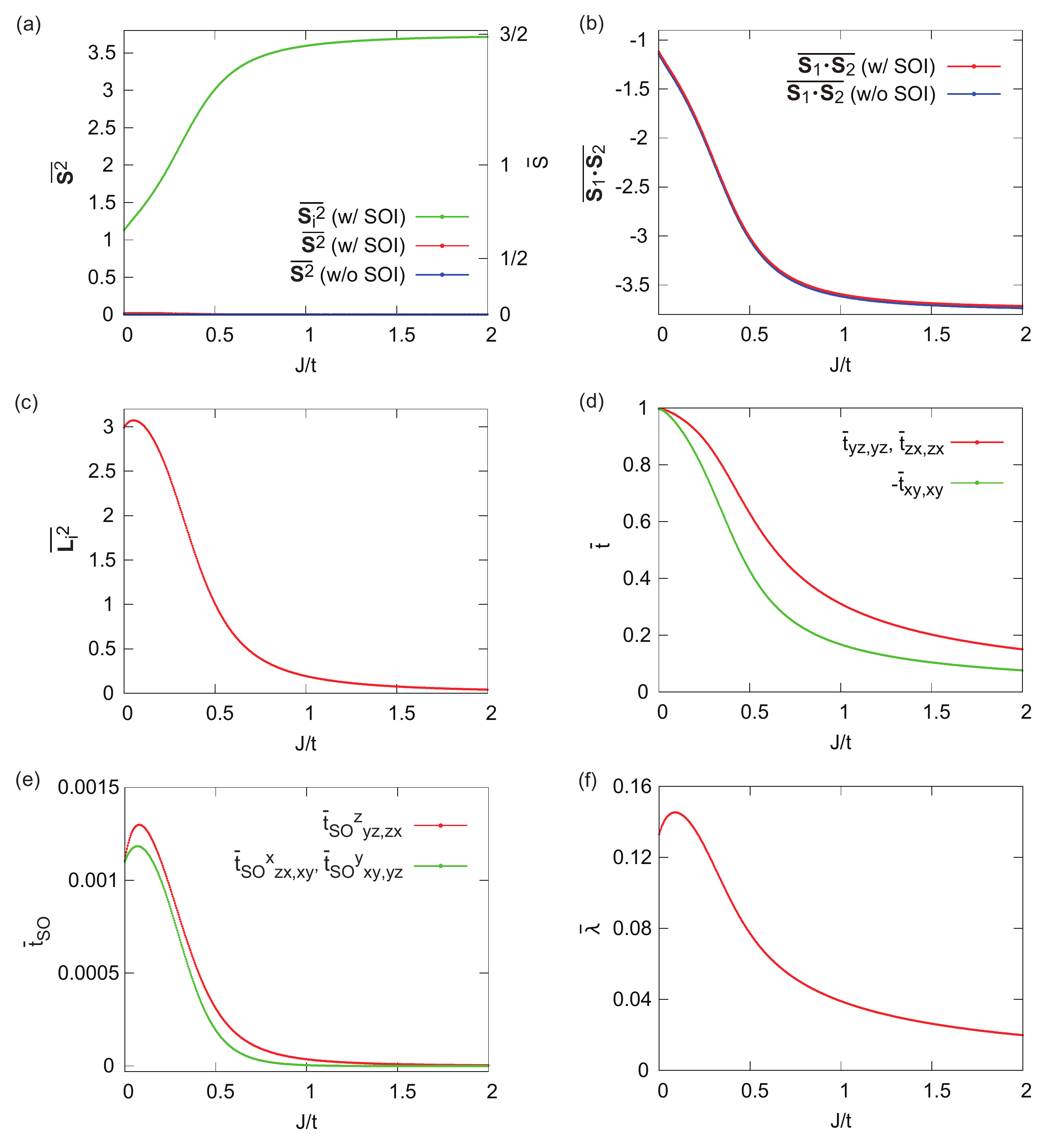}
\caption{(Color online) 
Results of exact diagonalization for the six-electron system: 
(a) Local spin angular momentum $\overline{\bm{S}_i^2}$ in the presence of SOI (green) and total spin angular momentum $\overline{\bm{S}^2}$ in the presence (red) and absence (blue) of SOI, 
(b) spin correlation $\overline{\bm{S}_1 \cdot \bm{S}_2}$ in the presence (red) and absence (blue) of SOI, 
(c) local orbital angular momentum $\overline{\bm{L}_i^2}$, 
(d) effective transfer $\bar{t}$'s, 
(e) effective interatomic SOI $\bar{t}_\text{SO}$'s, and 
(f) effective on-site SOI $\bar{\lambda}$. 
We can find no discontinuity in the expectation values. The half-filled system is similar to the quantum Heisenberg antiferromagnet, and it is antiferromagnetic for any value of $J$. 
The effective interatomic SOI $\bar{t}_\text{SO}$ is tiny, even though both $\bar{t}$ and $\bar{\lambda}$ are rather not small.
}
\label{fig:6e}
\end{figure}

The result of the half-filled system, i.e., the case of six electrons (Fig.~\ref{fig:6e}), is completely different from those of the cases with other electron numbers. 

First, the total spin angular momentum $\overline{\bm{S}^2}$ has tiny values for the entire range of $J/t$, whereas the local spin angular momentum $\overline{\bm{S}_i^2}$ grows as the Hund's coupling increases. 
The difference between $\overline{\bm{S}^2}$ and $\overline{\bm{S}_i^2}$ [Fig.~\ref{fig:6e}(a)] yields antiferromagnetic spin correlation [Fig.~\ref{fig:6e}(b)], following the relation Eq.~\eqref{eq:spin_cor}. 
It is a sharp contrast to the systems with less than six electrons, where they exhibit ferromagnetic correlation with the total spin $S$ maximized in the strong Hund's coupling region. 
This is reasonable since the effective Hamiltonian in the strong coupling limit is the quantum antiferromagnetic Heisenberg model in the half-filling case. 
Away from the half-filling, even a single hole drastically alters the ground state of the system, which is similar to the $t$-$J$ model. 

Another issue is the significant difference between $\bar{t}_\text{SO}$ and $\bar{\lambda}$ as shown in Figs.~\ref{fig:6e}(e) and \ref{fig:6e}(f). 
It results from the difference between the spin $S_i$ on each site and the total spin $S$ of the two sites. 
The effective intersite SOI $\bar{t}_\text{SO}$ reflects the total spin $S$. 
On the other hand, the effective on-site SOI $\bar{\lambda}_i$ has the information of the local $S_i$. 

We should also note the nonzero value of the orbital angular momentum in the weak coupling region [Fig.~\ref{fig:6e}(b)]. 
One might expect that the orbital angular momentum $\overline{\bm{L}_i^2}$ is quenched at each site 
because three electrons occupy the three $t_{2g}$ orbitals.  
However the itinerancy of electrons makes the expectation value of the orbital angular momentum $\bar{L}_i$ nonzero. 
In the strong coupling limit, the electron transfer is reduced by the Mott physics, which results in the quenched orbital degrees of freedom $L_i$. 
It leads to the smaller effective SOI's with increasing $J/t$.

\section{Discussion}

We have considered the two-site model of $t_{2g}$ orbitals, and have extended the notion of Hund's rule for two-atom systems. 
For the two-site model, we have observed completely different behaviors expected for a single-atom model by the Hund's rule due to the presence of electron transfer. 
The Hund's rule for a single atom is modified for the two-atom system as follows.

First, the spin angular momentum $S$ depends on the ratio of the Hund's coupling $J$ 
and the transfer energy $t$.  When the electron transfer is dominant, electrons are itinerant with a small spin $S$. 
As the ratio $J/t$ increases, Hund's rule of a single atom 
becomes relevant, and the spin $S$ is maximized 
in the large-$J/t$ limit. 
In detail, there are two kinds of spin values, namely the total spin $S$ 
of the two-site system and the local 
spin $S_i$, i.e., the spin angular momentum at site $i$. 
The total spin $S$ grows as $J/t$ increases except for the half-filling case, where the system 
consists mostly of $S=0$ states. 
On the other hand, the spin of a single atom $S_i$ always increases as $J/t$ becomes larger including the half-filling case. 
The total spin $\bar{S}$ and local spin $\overline{\bm{S}_i^2}$ in the large-$J$ limit are
\begin{gather}
\bar{S}  =  
\begin{cases}
\dfrac{n}{2} & (n \leq 5), \\
0 & (n=6), 
\end{cases} \\
\overline{\bm{S}_i^2} = \frac{1}{2} \left[ \frac{n_{+}}{2} \left( \frac{n_{+}}{2}+1 \right) + \frac{n_{-}}{2} 
\left( \frac{n_{-}}{2}+1 \right) \right],
\end{gather}
where $n_{+}$ and $n_{-}$ are given by $n_{+}=n$, $n_{-}=0$ for $n \leq 3$ and $n_{+}=3$, $n_{-}=n-3$ for $4 \leq n \leq 6$. 
The difference between $\bar{S}$ and $\bar{S}_i$ is manifested in the spin correlation $\overline{\bm{S}_1 \cdot \bm{S}_2}$ 
[see Eq.~\eqref{eq:spin_cor}]. 
Especially for the half-filling (six electrons) system, the total spin $S=0$ gives the antiferromagnetic correlation. 

Second, as for the orbital angular momentum $L_i$, it is maximized under the condition that the rule for spin we see above is satisfied. 
We have observed that except for the two-electron case $L_i$ gets smaller as the Hund's coupling $J$ increases, which is in contrast to the spin $S$. 
This means that the effect of the transfer $t$ is to increase the orbital angular momentum $L_i$ at each atom. 
As the Hund's coupling $J$ increases, electrons of parallel spin spread over three $t_{2g}$ orbitals to avoid the energy loss by electron correlation, especially on-site Hubbard $U$. 
It constrains and quenches the orbital degree of freedom. 
Thus the orbital angular momentum $L_i$ becomes smaller as the Hund's coupling $J$ and simultaneously $U(>U')$ increases. 

Hund's third rule is on the SOI. 
For the two-site model there are two kinds of effective SOI's: one is the on-site SOI $\bar{\lambda}_i$ and the other is the intersite SOI $\bar{t}_\text{SO}$. 
The magnitude of the effective SOI's is determined by both the spin and orbital angular momenta. 
We saw that the effective SOI's are largest in the region where both the spin and orbital degrees have moderate values, i.e., the intermediate-spin region. 
The effective SOI's become larger by electron correlation in the cases of four and five electrons. 

For the case of six electrons (half-filling), we need to be careful about the difference between the local spin $\bar{S}_i$ and total spin $\bar{S}$. 
The difference between the local spin $\bar{S}_i$ and total spin $\bar{S}$ also manifests itself as the difference between $\bar{\lambda}_i$ and $\bar{t}_\text{SO}$. 
Since the total spin $\bar{S}$ is vanishing, the effective intersite SOI $\bar{t}_\text{SO}$ is tiny for all ranges of $J$, in contrast to the local spin $\bar{S}_i$ and the effective on-site SOI $\bar{\lambda}_i$. 

Our findings would be a useful guideline to find a material that realizes effectively strong SOI.
They are summarized as follows. 
(i) The Hund's coupling $J$ and the electron number are the essential parameters for the 
enhancement of the interatomic SOI $\bar{t}_\text{SO}$ and the on-site SOI $\bar{\lambda}$. 
(ii) In the two-and three-electron systems, $J$ always suppresses $\bar{t}_\text{SO}$ and $\bar{\lambda}$ 
by increasing the ferromagnetic correlation.
(iii) In the four-and five-electron systems, the intermediate $J$ corresponding to the intermediate 
spin $S$ enhances $\bar{t}_\text{SO}$ and $\bar{\lambda}$. This indicates that the situation of the spin frustration or 
fluctuation is preferred by the enhanced effective SOI.
(iv) In the six-electron system, i.e., at half-filling, $\bar{t}_\text{SO}$ is always very small  
while $\bar{\lambda}$ is not, and both are suppressed 
by $J$ with the increase of antiferromagnetic correlation.

The candidate magnetic ions for the scenario of the enhanced effective SOI by electron correlation proposed in the present paper should have 
$t_{2g}^1$, $t_{2g}^2$, $t_{2g}^3$ configurations. 
In $3d$ elements, Sc$^{2+}$, Ti$^{3+}$, V$^{4+}$ have $t_{2g}^1$; Ti$^{2+}$, V$^{3+}$, Cr$^{4+}$ have $t_{2g}^2$; 
and V$^{2+}$, Cr$^{3+}$, Mn$^{4+}$ have $t_{2g}^3$ configuration. 
Considering perovskite and layered-perovskite transition metal oxides, where $t_{2g}$ and $e_g$ 
splitting  occurs due to the oxygen crystal field, 
the following materials are of particular interest. 
First, LaVO$_3$ is a Mott insulator consisting of V$^{3+}$ 
\cite{tokura1994mip,arima1993vog,miyasaka2002amh,nakamura1979spp}. 
This material has been studied in the context of metal-insulator transition due to electron correlation. 
From the viewpoint of the present paper, the effective SOI of this material could be enhanced by 
electron correlation with $t_{2g}^2$ configuration.  
Other candidate materials that would exhibit the enhanced effective SOI are 
La$_x$Sr$_{1-x}$CrO$_3$ ($0 \leq x \leq 0.5$) \cite{arevalo-lopez2008eel,chamberland1967pap,peck1999pds} 
with $t_{2g}^{2-2.5}$ configuration, 
LaSrVO$_4$ \cite{longo1973sla} with $t_{2g}^2$ configuration, and 
(La$_x$Sr$_{1-x}$)$_2$CrO$_4$ ($0 \leq x \leq 0.5$) \cite{peck1999pds} with $t_{2g}^{2-2.5}$ configuration. 

The present work extends the Hund's rule to systems with multiple atoms and itinerant electrons. 
Also, there is a case that the SOI is effectively enhanced by electron correlation, and it paves the way 
to design the materials having the effectively strong SOI. 
It would be possible by tuning the electron number and Hund's coupling even without using heavy elements. 

\begin{acknowledgments}
The authors acknowledge the fruitful discussion with Satoshi Okamoto. 
H.I. is supported by Grant-in-Aid for JSPS Fellows. 
This work is supported by Grant-in-Aid for Scientific Research (S)
(Grant No.~24224009) from the Ministry of Education, Culture,
Sports, Science and Technology of Japan, Strategic
International Cooperative Program (Joint Research Type)
from Japan Science and Technology Agency, and Funding
Program for World-Leading Innovative RD on Science and
Technology (FIRST Program).
\end{acknowledgments}

\begin{appendix}
\section{Symmetry}
\label{sec:symmetry}

\begin{figure}
\centering
\includegraphics[width=0.8\hsize]{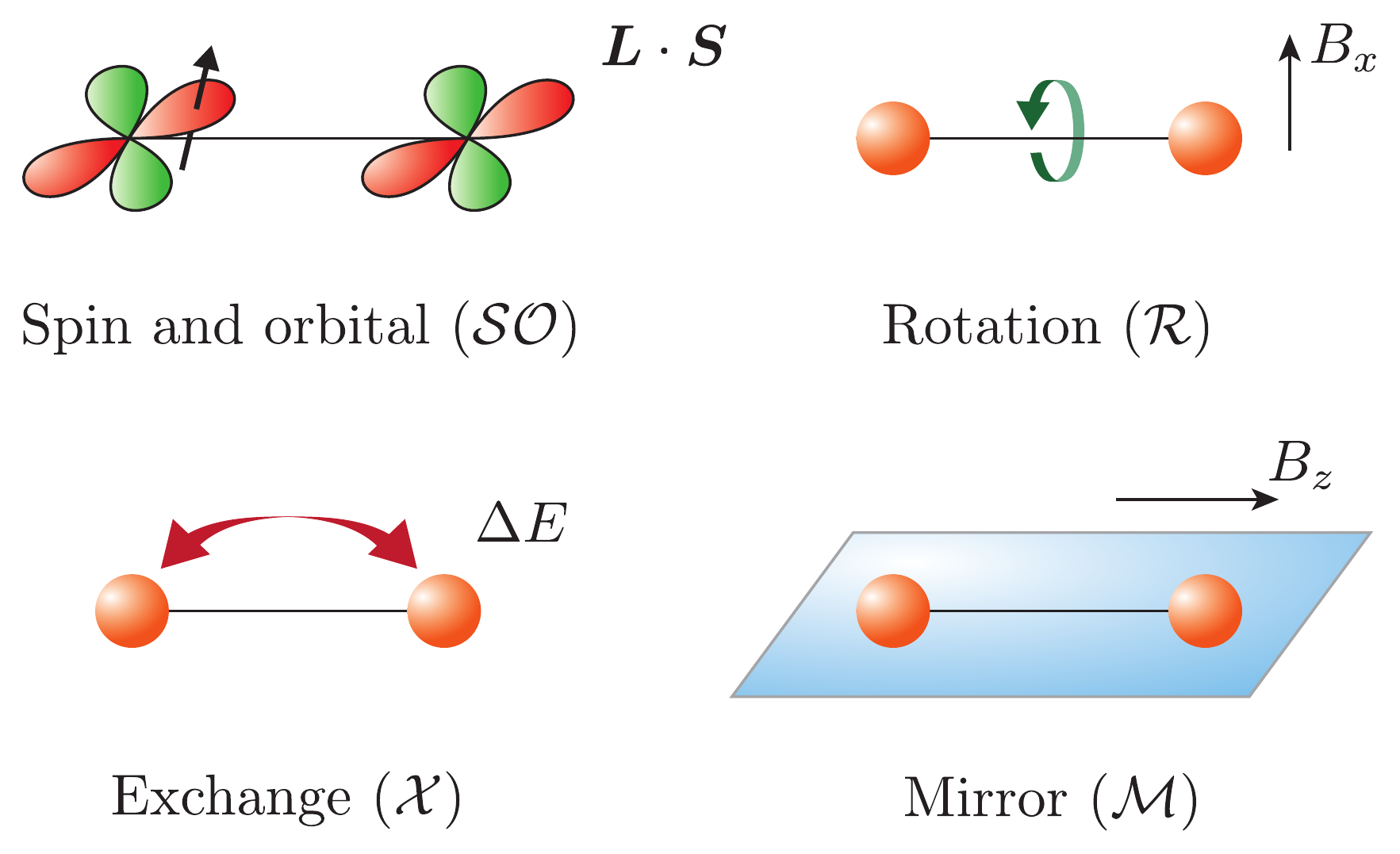}
\caption{(Color online) Symmetries of the two-site system and corresponding symmetry-breaking perturbation.}
\label{fig:symmetry}
\end{figure}

\begin{table}[b]
\caption{Changes of symmetry with corresponding discontinuities among the low-spin (L), intermediate-spin (I), and high-spin (H) states. 
For the four-electron system in the presence of SOI, the high-spin region consists of three areas, Ha, Hb, and Hc. 
The discontinuities and the corresponding symmetries are dependent of the number of electrons in the system $n$. 
}
\label{table}
\begin{tabular}{clllll}
\hline \hline
 & \multicolumn{5}{c}{Discontinuity} \\
\cline{2-6}
\ \ $n$ \ \  & L -- H & L -- I & I -- H (Ha) & Ha -- Hb & Hb -- Hc \\
\hline
2 & $\mathcal{SO}$ & --- &--- & --- & --- \\
3 & $\mathcal{SO}$, $\mathcal{X}$ & --- & \centering --- & --- & --- \\
4 & --- & $\mathcal{SO}$, $\mathcal{X}$, $\mathcal{M}$ \ & $\mathcal{SO}$, $\mathcal{R}$ & $\mathcal{SO}$, $\mathcal{R}$ & $\mathcal{SO}$, $\mathcal{R}$ \\
5 & --- & $\mathcal{SO}$ & $\mathcal{SO}$, $\mathcal{X}$, $\mathcal{M}$, $\mathcal{R}$ \ & --- & --- \\
\hline \hline
\end{tabular}
\end{table}

The simple model we consider here has four symmetries; i.e., 
the spin and orbital symmetry ($\mathcal{SO}$), the rotation around the bond of the two atoms ($\mathcal{R}$), 
the exchange of the two atoms ($\mathcal{X}$), and the mirror symmetry in a plane containing the two atoms ($\mathcal{M}$) 
(Fig.~\ref{fig:symmetry}). 
Now we assume that the bond of the two atoms directs the $z$ axis, and then one of the mirror symmetry plane 
will be the $yz$ plane. 
These four symmetries separate the Hamiltonian into some blocks, and it leads to a discontinuity of expectation values for the ground states.
If a discontinuity exists, it will be lifted by introducing the proper symmetry-breaking term.
The spin and orbital symmetry can be broken by the SOI, 
the rotation symmetry by the magnetic field perpendicular to the bond (e.g., $B_x$), 
the exchange symmetry by the energy level difference between the atoms ($\Delta E$), and
the mirror symmetry by the magnetic field parallel to the bond ($B_z$). 
The energy under the magnetic field is given by the Zeeman energy, which is given by
\begin{equation}
\hat{H}_\text{B} = -\frac{e}{2mc} \bm{B} \cdot ( \hat{\bm{L}} + 2\hat{\bm{S}} ).
\end{equation}

We start from a highly symmetric system to find what symmetry is related to the discontinuity of order parameters; 
namely we consider the Hamiltonian 
\begin{equation}
\hat{H} = \hat{H}^\text{(t)}_{12} + \sum_{i=1}^2 \hat{H}^\text{(Kanamori)}_i
\end{equation}
as a starting point. 
There are no SOI, level difference, and magnetic field. 
Discontinuities appear with the change of the total spin $S$. 
The discontinuities and the corresponding symmetry-breaking terms are summarized in Table~\ref{table}. 
Especially in the four-electron system, the SOI separates the high-spin region into three parts, which merge again by breaking $\mathcal{R}$.

\end{appendix}

\bibliography{jabref}

\end{document}